\RequirePackage{lineno}
\documentclass[twocolumn,showpacs,superscriptaddress,amsmath,amssymb]{revtex4-1}
\usepackage{graphicx}
\usepackage{epsfig}
\usepackage{dcolumn}
\usepackage{bm}
\usepackage{overpic}
\usepackage{color}
\usepackage{lineno}
\usepackage{multirow}
\usepackage{subfigure}
\begin{document}
\title{\boldmath Complete measurement of $\Lambda$ electromagnetic form factors}
\author{
\begin{small}
M.~Ablikim$^{1}$, M.~N.~Achasov$^{10,d}$, P.~Adlarson$^{59}$, S. ~Ahmed$^{15}$, M.~Albrecht$^{4}$, M.~Alekseev$^{58A,58C}$, A.~Amoroso$^{58A,58C}$, F.~F.~An$^{1}$, Q.~An$^{55,43}$, Y.~Bai$^{42}$
% ...
\\
\vspace{0.2cm}
(BESIII Collaboration)\\
\vspace{0.2cm} {\it
$^{1}$ Institute of High Energy Physics, Beijing 100049, People's Republic of China\\
% ...
$^{4}$ Bochum Ruhr-University, D-44780 Bochum, Germany\\
% ...
$^{10}$ G.I. Budker Institute of Nuclear Physics SB RAS (BINP), Novosibirsk 630090, Russia\\
% ...
$^{15}$ Helmholtz Institute Mainz, Johann-Joachim-Becher-Weg 45, D-55099 Mainz, Germany\\
% ...
$^{42}$ Southeast University, Nanjing 211100, People's Republic of China\\
$^{43}$ State Key Laboratory of Particle Detection and Electronics, Beijing 100049, Hefei 230026, People's Republic of China\\
% ...
$^{55}$ University of Science and Technology of China, Hefei 230026, People's Republic of China\\
% ...
$^{58}$ (A)University of Turin, I-10125, Turin, Italy; (B)University of Eastern Piedmont, I-15121, Alessandria, Italy; (C)INFN, I-10125, Turin, Italy\\
$^{59}$ Uppsala University, Box 516, SE-75120 Uppsala, Sweden\\
% ...
$^{d}$ Also at the Novosibirsk State University, Novosibirsk, 630090, Russia\\
% ...
}
\vspace{0.4cm}
}
\noaffiliation{}
  
\date{\today}

\begin{abstract}
The exclusive process $e^+e^-\rightarrow\Lambda\bar{\Lambda}$, with $\Lambda \to p\pi^-$ and $\bar{\Lambda} \to \bar{p}\pi^+$, has been studied at $\sqrt{s} =$ 2.396 GeV for measurement of the $\Lambda$ electric and magnetic form factors, $G_E$ and $G_M$. A data sample, corresponding to an integrated luminosity of 66.9 pb$^{-1}$, was collected with the BESIII detector for this purpose. A multi-dimensional analysis with a complete decomposition of the spin structure of the reaction enables a determination of the modulus of the ratio $R=|G_E/G_M|$ and, for the first time for any baryon, the relative phase $\Delta\Phi=\Phi_E-\Phi_M$. The resulting values are obtained using the recent and most precise measured value of the asymmetry parameter $\alpha_{\Lambda}$ = $0.750~\pm~0.010$ to be $R~=~0.96\pm0.14~(\rm stat.)\pm~0.02~(\rm sys.)$ and $\Delta\Phi=37^{\mathrm{o}}\pm~12^{\mathrm{o}}~(\rm stat.)\pm~6^{\mathrm{o}}~(\rm sys.)$, respectively. In addition, the cross section is measured with unprecedented precision to be $\sigma = 118.7~\pm~5.3~(\rm stat.)\pm~5.1~(\rm sys.)$ pb, which corresponds to an effective form factor of $|G|=0.123~\pm~0.003~(\rm stat.)\pm~0.003~(\rm sys.)$. The contribution from two-photon exchange is found to be negligible. Our result enables the first complete determination of baryon time-like electromagnetic form factors.
\end{abstract}

\maketitle
\noindent One of the most challenging questions in contemporary physics is to understand the strong interaction in the confinement domain, \textit{i.e.} where quarks form hadrons. This puzzle manifests itself in one of the most abundant building blocks of the Universe: the nucleon. Despite being known for a century, we still do not understand its size~\cite{pradius}, its spin~\cite{pspin}, nor its intrinsic structure~\cite{pstructure}. The latter has been extracted from space-like electromagnetic form factors (EMFFs), fundamental properties of hadrons that have been studied since the 1960's~\cite{ffreview}. In particular, the neutron charge distribution is very intriguing~\cite{pstructure}. Hyperons provide a new angle on the nucleon puzzle: What happens if we replace one of the $u$- and $d$-quarks with a heavier $s$-quark? A systematic comparison of octet baryons sheds light on to what extent SU(3) flavour symmetry is broken. The importance of hyperon structure was pointed out as early as 1960~\cite{CabbibboGatto}, but has not been objected to rigorous experimental studies until now. The main reason is that space-like EMFFs of hyperons are not straight-forward to access experimentally since their finite life-time make them unsuitable as beams and targets. However, the recent development of high-intensity electron-positron colliders in the strange- and charm energy region offers a viable approach to the quantization of hyperon structure in the \textit{time-like} region. 

Spin 1/2 baryons are described using two independent EMFFs, commonly the electric form factor $G_E$ and the magnetic form factor $G_M$. These can be studied in $e^+e^-\rightarrow B\bar{B}$ reactions and are functions of the four-momentum transfer $s=q^2$: $G_E \equiv G_E(s)$ and $G_M \equiv G_M(s)$. In the time-like region, where $s$ is positive, EMFFs can be complex with a relative phase~\cite{theorypolarization}. This phase, $\Delta\Phi \equiv \Delta\Phi(s)$, is a result of interfering amplitudes corresponding to different partial waves. Hence it must be zero at the kinematic threshold, where only the $s$-wave contributes. Furthermore, analyticity requires that the phase goes to zero as $s \to \infty$, since space-like and time-like EMFFs should converge to the same value. However, for intermediate $s$ the phase can be non-zero. This would introduce polarization effects on the final state, even if the initial state is unpolarized~\cite{theorypolarization}. Thanks to the weak, parity violating decays of hyperons, the polarization is experimentally accessible. This provides unique opportunities compared to nucleons.

The first measurement of $e^+e^- \to \Lambda \bar{\Lambda}$ production was reported by the DM2 collaboration~\cite{DM2}. The first determination of the $\Lambda$ EMFFs was provided by the BaBar collaboration, using the initial state radiation (ISR) method~\cite{BaBarll}. However, the sample was insufficient for a clear separation of the electric and magnetic form factors. An attempt was made to extract the phase from the $\Lambda$ polarization, but the result was inconclusive~\cite{BaBarll}. The cross section of $e^+e^-$ production of protons and ground-state hyperons at $\sqrt{s}$ = 3.69, 3.77 and 4.17 GeV was measured with CLEO-c data. The magnetic form factors were extracted assuming $|G_E|$ = $|G_M|$~\cite{CLEOc}. The \mbox{BESIII} collaboration performed in 2011-2012 an energy scan, enabling an investigation of the $\Lambda$ production cross section at four energies between $\sqrt{s}=2.23$ and $\sqrt{s}=3.08$ GeV. An unexpected enhancement at the kinematic threshold was observed~\cite{Xiaorong}. At higher energies, the statistical precision was improved compared to previous experiments, though still not sufficient to extract the form factor ratio $R\equiv\vert G_E / G_M \vert$. The recent experimental progress has resulted in an increasing interest from the theory community. For instance, predictions of the relative phase have been made, based on various $\Lambda\bar{\Lambda}$ potential models~\cite{JU} with input data from the PS185 experiment~\cite{PS185}.

In this Letter, the exclusive process $e^+e^-\to\Lambda\bar{\Lambda}$ ($\Lambda\to p\pi^-$, $\bar{\Lambda}\to\bar{p}\pi^+)$ is studied at $\sqrt{s}$ = 2.396~GeV. In the following, we present our measurements of the cross section $\sigma \equiv \sigma(s)$, the ratio $R=\vert G_E / G_M \vert$ and, for the first time, the relative phase $\Delta\Phi$.

Assuming one-photon exchange ($e^+e^-\rightarrow\gamma^*\rightarrow B\bar{B}$), the \textit{Born cross section} of spin 1/2 baryon-antibaryon pair production can be parameterized in terms of $G_E$ and $G_M$:
\begin{equation}
\sigma_{B\bar{B}}(s) = \frac{4\pi\alpha^2\beta}{3s}\left[|G_M(s)|^2+\frac{1}{2\tau}|G_E(s)|^{2}\right].
\label{equ-borncs}
\end{equation}

\noindent Here, $\alpha$=1/137.036 is the fine-structure constant, $\beta=\sqrt{1-4m^2_B/s}$ the velocity of the produced baryon, $m_B$ the mass of the baryon, and $\tau = s/(4m^2_B)$. 

The \textit{effective form factor} is defined as
\begin{equation}
\begin{split}
|G(s)| \equiv &\sqrt{\frac{\sigma_{B\bar{B}}(s)}{(1+\frac{1}{2\tau})(\frac{4\pi\alpha^2\beta }{3s})}} \\
\equiv &\sqrt{\frac{2\tau|G_M(s)|^2+|G_E(s)|^2}{2\tau+1}}.
\label{equ-effectiveff}
\end{split}
\end{equation}

A complete decomposition of the complex $G_E$ and $G_M$ requires a multi-dimensional analysis of the reaction and the subsequent decays of the produced baryons. In Refs.~\cite{Faldt:2016qee,Faldt:2017kgy}, the joint decay distribution of $e^+e^-\to\Lambda\bar{\Lambda} (\Lambda\to p\pi^-, \bar{\Lambda}\to\bar{p}\pi^+)$ was derived in terms of the phase $\Delta\Phi$ and the angular distribution parameter $\eta=(\tau-R^2)/(\tau+R^2)$:
\begin{equation}
\label{angdist}
\begin{split}
{\cal{W}}({\boldsymbol{\xi}})=&{\cal{T}}_0+{{{\eta}}}{\cal{T}}_5\\
-&{{\alpha_\Lambda^2}}\left({\cal{T}}_1
+\sqrt{1-{{\eta}}^2}\cos({{\Delta\Phi}}){\cal{T}}_2
+{{\eta}}{\cal{T}}_6\right)\\
+&{\alpha_\Lambda}\sqrt{1-{{\eta}}^2}\sin({{\Delta\Phi}})
\left({\cal{T}}_3-{\cal{T}}_4\right),
\end{split}
\end{equation}
where $\alpha_{\Lambda}^{PDG}$ denotes the decay asymmetry of the $\Lambda\to p\pi^-$ decay. The seven functions ${\cal{T}}_k({\boldsymbol{\xi}})$ do not depend on the physical quantities $\eta$ and $\Delta\Phi$, but only on the measured angles:
\begin{align}
	{\cal{T}}_0({\boldsymbol{\xi}}) =&1,\nonumber\\
	{\cal{T}}_1({\boldsymbol{\xi}}) =&{\sin^2\!\theta}\sin\theta_1\sin\theta_2\cos\phi_1\cos\phi_2+
{\cos^2\!\theta}\cos\theta_1\cos\theta_2,\nonumber\\
	{\cal{T}}_2({\boldsymbol{\xi}}) =&{\sin\theta\cos\theta}\left(\sin\theta_1\cos\theta_2\cos\phi_1+
\cos\theta_1\sin\theta_2\cos\phi_2\right),\nonumber\\
	{\cal{T}}_3({\boldsymbol{\xi}}) =&{\sin\theta\cos\theta}\sin\theta_1\sin\phi_1,\nonumber\\
	{\cal{T}}_4({\boldsymbol{\xi}}) =&{\sin\theta\cos\theta}\sin\theta_2\sin\phi_2,\nonumber\\
    {\cal{T}}_5({\boldsymbol{\xi}}) =&{\cos^2\!\theta},\nonumber\\
	{\cal{T}}_6({\boldsymbol{\xi}}) =&{\cos\theta_1\cos\theta_2
    -\sin^2\theta\sin\theta_1\sin\theta_2\sin\phi_1\sin\phi_2}.\nonumber
        \label{eq:tangles}
\end{align}
The five angles measured are: $\theta$, the $\Lambda$ scattering angle with respect to the electron beam; $\theta_1$ and $\phi_1$, the proton helicity angles from the $\Lambda\to p\pi^-$ decay; and $\theta_2$ and $\phi_2$, the antiproton helicity angles from the $\bar{\Lambda}\to\bar{p}\pi^+$ decay. The decay angles are defined in the rest system of the $\Lambda$ and the $\bar{\Lambda}$, respectively. We define a right-handed system where the $z$-axis is oriented along the $\Lambda$ momentum $\mathbf{p}_{\Lambda}=-\mathbf{p}_{\bar{\Lambda}}$ in the $e^+e^-$ rest system. The $y$-axis is perpendicular to the reaction plane and is oriented along the $\mathbf{k}_{e^-}\times\mathbf{p}_{\Lambda}$ direction, where $\mathbf{k}_{e^-}=-\mathbf{k}_{e^+}$ is the electron beam momentum in the $e^+e^-$ rest system. The definitions of the angles are illustrated in Fig.~\ref{fig:system}. 

The term ${\cal{T}}_0 + \eta {\cal{T}}_5$ in Eq.~\eqref{angdist} describes the scattering angle distribution of the $\Lambda$ hyperon. The term ${\alpha_\Lambda}\sqrt{1-{{\eta}}^2}\sin({{\Delta\Phi}})({\cal{T}}_3-{\cal{T}}_4)$ accounts for the transverse polarization $P_y$ of the $\Lambda$ and $\bar{\Lambda}$. 

In particular, the $\Lambda$ transverse polarization $P_y$ is given by:
\begin{equation}
P_y=\frac{\sqrt{1-\eta^2}\sin\theta\cos\theta}{1+\eta\cos^2\theta}\sin(\Delta\Phi).
\end{equation}
Finally, the  ${{\alpha_\Lambda^2}}({\cal{T}}_1+\sqrt{1-{{\eta}}^2}\cos({{\Delta\Phi}} ){\cal{T}}_2+{{\eta}}{\cal{T}}_6)$ term describes the spin correlations between the two hyperons.
 
The asymmetry parameter $\alpha_{\Lambda}^{PDG}$ is $0.642~\pm~0.013$ according to PDG ~\cite{PDG}. However, a recent measurement of $J/\psi\to\Lambda\bar{\Lambda}$ by the BESIII collaboration~\cite{BAM116} revealed a significantly different value of the decay asymmetry parameter of $\alpha_{\Lambda}~=~0.750~\pm~0.010$. In our opinion
the  BESIII value is preferred over $\alpha_{\Lambda}^{PDG}$ 
which was established more than forty years ago and there are strong indications that not all systematic effects were considered.   

\begin{figure}[htbp]
\setlength{\abovecaptionskip}{-0.5cm}
\setlength{\belowcaptionskip}{-0.5cm}
\begin{center}
\includegraphics[width=3.14in,height=2.22in,angle=0]{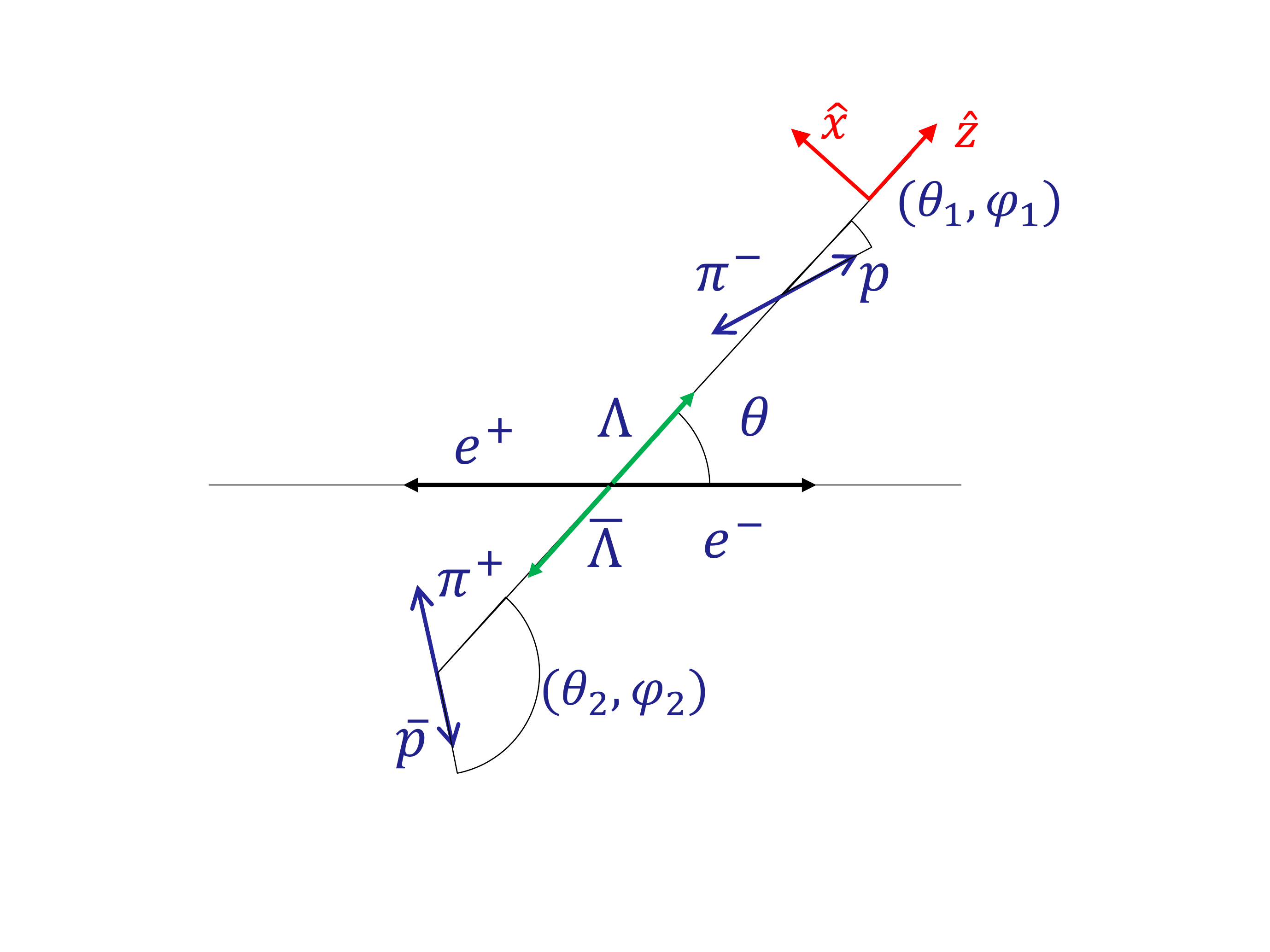}
\caption{Definition of the coordinate system used to describe the $e^+e^-\to\Lambda\bar{\Lambda}$ ($\Lambda\to p\pi^-$, $\bar{\Lambda}\to\bar{p}\pi^+)$ process. }
\label{fig:system}
\end{center}
\end{figure}

A data sample corresponding to an integrated luminosity of 66.9 pb$^{-1}$ was collected with the Beijing Spectrometer (BESIII) at the Beijing Electron Positron Collider (BEPCII). The BESIII detector has a geometrical acceptance of 93\% of the solid angle. BESIII contains a small-cell, helium-based main drift chamber (MDC), a time-of-flight system (TOF) based on plastic scintillators, an electromagnetic calorimeter (EMC) made of CsI(Tl) crystals, a muon counter (MUC) made of resistive plate chambers, and a superconducting solenoid magnet with a central field of 1.0 Tesla. A detailed description of the detector and its performance can be found in Ref.~\cite{BESIII}.

The particle propagation through the detector is modeled using a {\sc Geant}-based~\cite{geant4} Monte Carlo (MC) simulation software package, {\sc Boost}~\cite{boost}. The multi-dimensional analysis for determination of $R$ and $\Delta\Phi$ enables a model-independent efficiency correction. The simulations for this purpose are performed with a MC sample generated by a phase space generator. The final simulations of $e^+e^-\to\Lambda\bar{\Lambda}$ ($\Lambda\to p\pi^-,\bar{\Lambda}\to\bar{p}\pi^+$) for determination of $\sigma$ and $G$ are performed with the measured values of $G_E/G_M$ as input to the {\sc ConExc} generator~\cite{conexc}. In {\sc ConExc}, high-order processes with one radiative photon are taken into account. For background studies, an inclusive MC sample of continuum processes $e^+e^-\to q\bar{q}$ with $q=u, d, s$ is used. 

In the analysis, events are reconstructed by the final state particles $p$, $\pi^-$, $\bar{p}$ and $\pi^+$. We therefore require at least four charged tracks per event. Each track must be reconstructed within the MDC, \textit{i.e} with polar angles $\theta$ fulfilling $|\cos\theta|<$0.93, measured in the laboratory frame between the direction of the track and the direction of the $e^+$ beam. The momentum of each track must be smaller than 0.5 GeV/$c$. Based on simulations, we identify tracks with momenta less than 0.2 GeV/$c$ as $\pi^+/\pi^-$ candidates, whereas tracks with momenta larger than 0.2 GeV/$c$ are identified as $p/\bar{p}$ candidates. 

The $\Lambda$ and $\bar{\Lambda}$ candidates are reconstructed by fitting each $p\pi^-$ ($\bar{p}\pi^+$) to a common vertex corresponding to the decay of $\Lambda(\bar{\Lambda})$. A four-constraint (4C) kinematic fit is applied on the $\Lambda$ and $\bar{\Lambda}$ candidates, using energy and momentum conservation in $e^+e^- \to \Lambda \bar{\Lambda}$ and requiring $\chi^2_{\rm 4C}<50$. 
We require the $p\pi^-$/$\bar{p}\pi^+$ invariant mass to fulfill $|M(p\pi^-/\bar{p}\pi^+)-m_{\Lambda}|<$ 6 MeV/$c^2$. The $M(p\pi^-)$ distribution is shown in Fig.~\ref{fig:mlam}. Here, $m_{\Lambda}$ is the nominal mass of $\Lambda$ from the PDG~\cite{PDG}. The mass window corresponds to $\pm~4\sigma$ of $|M(p\pi^-/\bar{p}\pi^+)|$ mass resolution. After applying all event selection criteria, 555 event candidates remain in our data sample. 

\begin{figure}[!htbp]

\setlength{\belowcaptionskip}{-0.5cm}
\begin{center}
\includegraphics[width=3.14in,height=2.22in,angle=0]{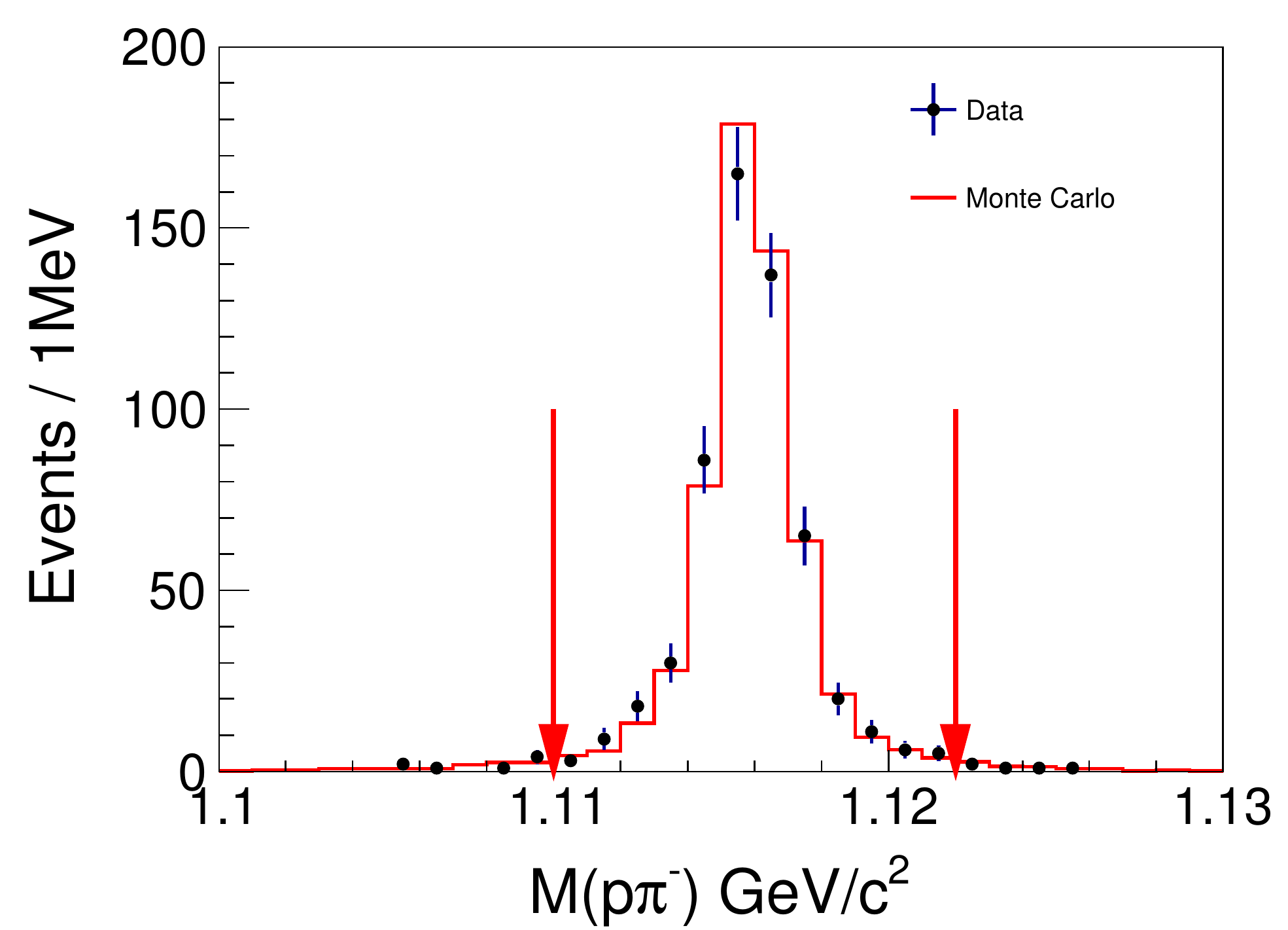}
\caption{The invariant mass of $p\pi^-$ for BESIII data (black dots) and Monte Carlo data (red line) fulfilling all selection criteria except those on invariant mass. The MC data are normalised to the total number of events in the data.}
\label{fig:mlam}
\end{center}
\end{figure}

The background channels are identified by performing inclusive $q\bar{q}$ simulations. The main contribution are events from $\Delta^{++}\bar{p}\pi^-(\bar{\Delta}^{--}p\pi^+)$ and non-resonant $p\bar{p}\pi^+\pi^-$ production, \textit{i.e.} reactions with similar topology as $e^+e^-\to\Lambda\bar{\Lambda}$ ($\Lambda\to p\pi^-$, $\bar{\Lambda}\to\bar{p}\pi^+$). The contamination is found to be on the percent level. A two-dimensional sideband study provides a data-driven method to quantify the background contribution. The $\Lambda$ sideband regions are defined within 1.097 GeV/$c^2$ $<M(p\pi^-/\bar{p}\pi^+)<$ 1.109 GeV/$c^2$ or 1.123 GeV/$c^2$ $<M(p\pi^-/\bar{p}\pi^+)<$ 1.135 GeV/$c^2$ for events with a $\bar{\Lambda}$ candidate. The $\bar{\Lambda}$ sidebands are defined in the corresponding way. The number of background events is determined to be $14\pm4$, corresponding to a background level of 2.5\%.

In our analysis, we extract the parameters $\eta$ and $\Delta\Phi$ by applying a multidimensional event-by-event maximum log-likelihood fit to our data. Using Eq.~\ref{angdist}, the probability of the $i$th event is given by:
\begin{equation}
{\cal{P}} (\boldsymbol{\xi}_i;\eta,\Delta\Phi) = {\cal{W}}(\boldsymbol{\xi}_i;\eta, \Delta\Phi)\epsilon({\boldsymbol{\xi}_i})/{\cal{N}}(\eta,\Delta \Phi),
\end{equation}
where  $\epsilon({\boldsymbol{\xi}_i})$ is the efficiency as a function of the scattering and decay angles, represented by the vector ${\xi}_i$. The normalization factor $\cal{N}$ is calculated for each parameter set using a sum of the corresponding $\cal{W({\boldsymbol{\xi}})}$ for phase space generated events and processed through detector simulation and reconstructed as the data sample. The joint probability density for $N$ events is
\begin{equation}
{\cal{L}} (\boldsymbol{\xi}_1, \boldsymbol{\xi}_2,..., \boldsymbol{\xi}_N ;\eta,\Delta\Phi) =
\prod_{i=1}^N\frac{{\cal{W}}(\boldsymbol{\xi}_i;\eta, \Delta\Phi)\epsilon({\boldsymbol{\xi}_i})}{{\cal{N}}(\eta,\Delta \Phi)}.
\label{eqn:linlik}
\end{equation}

The parameters $\eta$ and $\Delta\Phi$ are determined in {\sc MINUIT}~\cite{minuit} by minimizing the log-likelihood function:
\begin{equation}
-\ln{\cal{L}}=-\sum_{i=1}^{N}\ln \frac{{{\cal{W}}({\boldsymbol{\xi}_i};\eta,\Delta \Phi)}}{{\cal{N}}(\eta,\Delta \Phi)}-\sum_{i=1}^{N}
\ln\epsilon({\boldsymbol{\xi}_i})
\end{equation}
where the last term does not depend on the parameters $\eta$ and $\Delta\Phi$. For our nominal result we use the BESIII value of $\alpha_{\Lambda}$ in Eq.~\eqref{angdist}. The fit to the selected events results in $\eta=0.12\pm0.14$, 
giving $R~=~0.96~\pm~0.14$, and $\Delta\Phi~=~ 37^{\mathrm{o}}~\pm 12^{\mathrm{o}}$. The uncertainties are statistical only. The correlation coefficient between $\eta$ and $\Delta\Phi$ is 0.17.  The $\Lambda$ angular distribution and the polarization as a function of the scattering angle are shown in Fig.~\ref{cosThetaLambda}. 
If, instead, the PDG value for the $\Lambda$ decay parameter $\alpha_{\Lambda}^{PDG}$ is used, then $R$ becomes $0.94\pm0.16$ and the phase $\Delta\Phi$ = $42^{\mathrm{o}}\pm16^{\mathrm{o}}$. 
\begin{figure*}[!htbp]
\begin{center}
\includegraphics[width=0.49\textwidth,height=2.4in,angle=0]{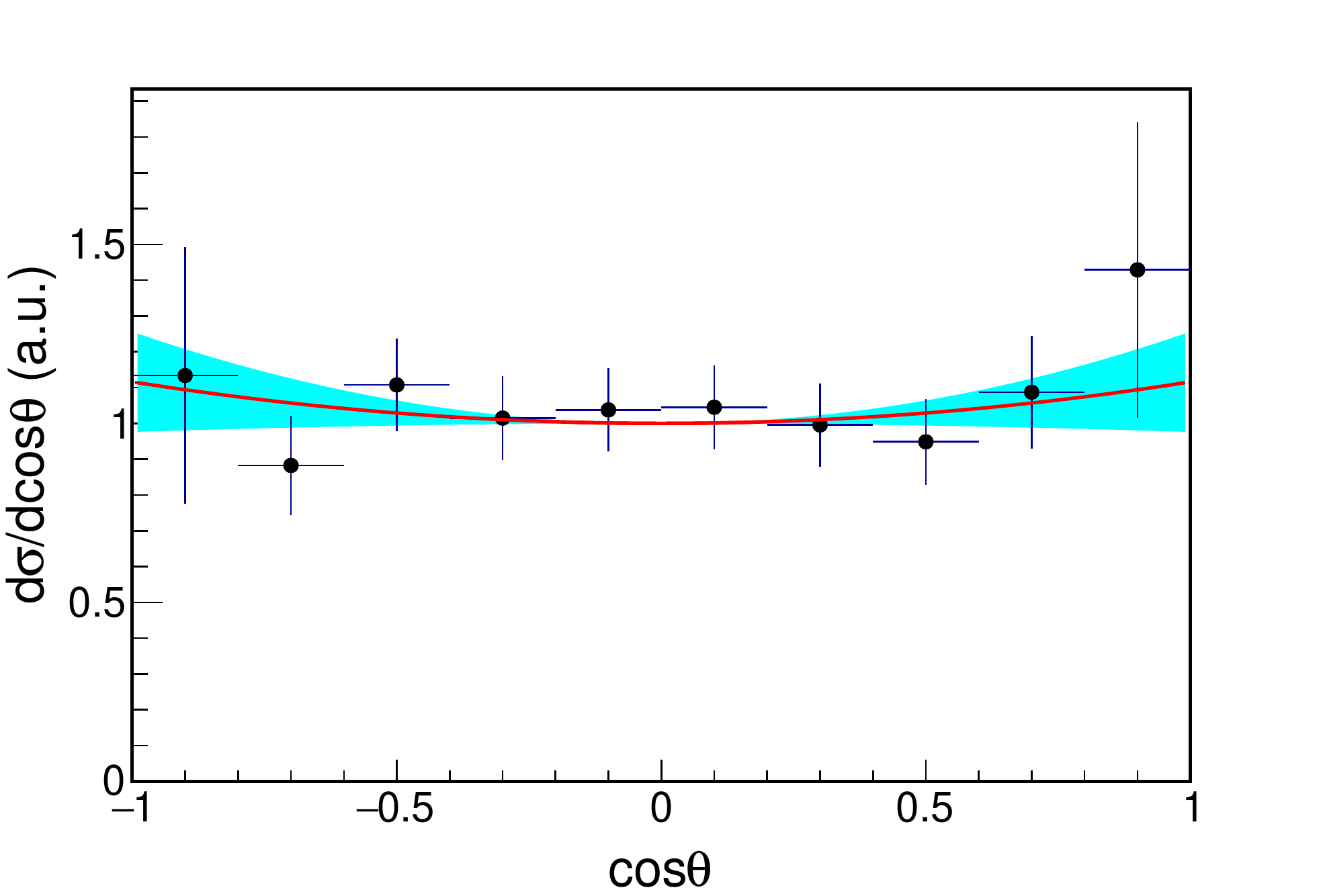}
\put(-45, 140){\large a)}
\includegraphics[width=0.5\textwidth,height=2.4in,angle=0]{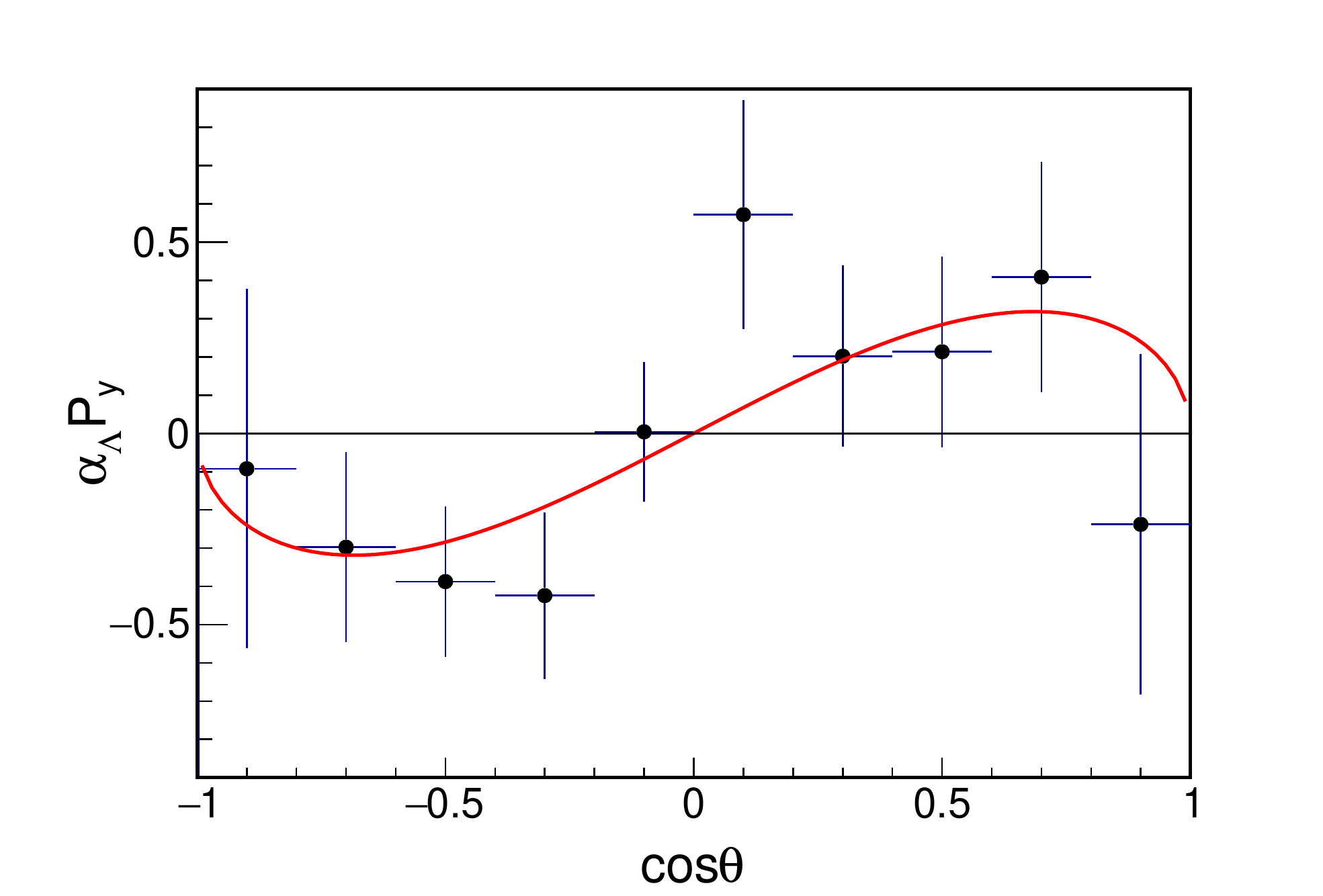}
\put(-45, 140){\large b)}
\caption{a) The acceptance corrected $\Lambda$ scattering angle distribution. The experimental distribution (points) is normalized to yield $A=1$  when fitting $A+B\cos^2\theta$ to the data. The red line is $1+\eta\cos^2\theta$ with $\eta=0.12$ and the band corresponds to the statistical
uncertainty.
 b) The product of $\alpha_\Lambda$ and $\Lambda$ polarization $P_y$ as a function of the scattering angle. The dots are the data, the red line the polarization corresponding to the $\Delta\Phi$ and $\eta$ obtained in the fit.
 The multidimensional acceptance correction to create the plots was obtained from the MC simulation using parameters determined from the maximum log-likelihood fit.}
\label{cosThetaLambda}
\end{center}
\end{figure*}

A thorough investigation of possible sources of systematic uncertainties has been performed. The
uncertainties from the luminosity measurement, tracking,
and background are found to be negligible. The non-negligible contributions from the angular fit range (for $R$), from requirements on $\chi^2_{\rm 4C}$ (for $\Delta\Phi$), and requirements on the invariant mass are summarized in Table I. The total systematic uncertainty is about seven times smaller than the statistical for $R$ and about two times smaller for $\Delta\Phi$.
\begin{table}[htbp]
\centering
\caption{Systematic uncertainties in $R$ and $\Delta\Phi$}
\label{sys}
\begin{tabular}{c | c | c  }
\hline
\hline
Source&$R$ (\%)&$\Delta\Phi$ (\%)\\\hline
$\chi^2_{\rm 4C}$ cut& - &14\\
mass window of $p\pi$&0.1&5.5\\
different range of $\cos\theta$& 2.0 & - \\
\hline
total&2.0&15 \\\hline\hline

\end{tabular}
\end{table}

The formalism presented in Eq.~\eqref{angdist} assumes the one-photon exchange to be dominant in the production mechanism. A significant contribution of two-photon exchange of the lowest order results in an
additional term $\kappa\cos\theta\sin^2\theta$ in Eq.~\eqref{angdist} due to interference of the one- and two-photon amplitudes~\cite{Gakh:2005wa}. This would give rise to a non-zero asymmetry
\begin{equation}
A=\frac{N(\cos\theta>0)-N(\cos\theta<0)}{N(\cos\theta>0)+N(\cos\theta<0)}
\end{equation}
in the $\Lambda$ angular distribution \cite{egle}. The asymmetry $A$ is related to $\kappa$ in the following way:
\begin{equation}
A=\frac{3}{4}\frac{\kappa}{3+\eta}.\label{eq:Abeta}
\end{equation}
In this work, the asymmetry is measured to be $A~=~0.001~\pm~0.037$ and indicates a negligible contribution from two-photon exchange with respect to the statistical precision.

The total cross section has been calculated using
\begin{equation}
\sigma_{\Lambda\bar{\Lambda}} = \frac{N_{\rm signal}}{\mathcal{L}_{\rm int}\epsilon(1+\delta)\mathcal{B}},
\label{equ-born}
\end{equation}
where $N_{\rm signal}=N_{\rm data}-N_{\rm bg}$, $N_{\rm data}$ = 555 is the number of events in the sample after all selection criteria, $N_{\rm bg}$ = 14$ \pm$ 4 the number of events in the sidebands, and $\mathcal{L}_{\rm int}$ the integrated luminosity. The reconstruction efficiency $\epsilon$ should in principle depend on the parameters $R$ and $\Delta\Phi$. However, simulations using the {\sc Phokhara} generator~\cite{czyz} show that the phase has negligible impact on the efficiency. In a recent measurement of the $\Lambda\bar\Lambda$ cross section at threshold by the BESIII collaboration~\cite{Xiaorong}, the largest source of systematics turned out to be the model dependence from $R$. In this work, we were able to minimize the systematics by measuring $R$ and evaluating the efficiency using a MC sample from the {\sc ConExc} generator with the measured $R$ as input. The radiative correction factor $1+\delta$ is determined taking ISR and vacuum polarization into account. The factor $\mathcal{B}$ is the product of the branching fractions of $\Lambda\to p\pi^-$ and $\bar{\Lambda}\to\bar{p}\pi^+$, taken from Ref.~\cite{PDG}.  

The following systematic effects contribute to the uncertainty of the cross section measurement: i) The uncertainty from the $\Lambda$ and $\bar{\Lambda}$ reconstruction is determined to be 1.1\% and 2.4\%, respectively, using single-tag samples of $\Lambda$ and $\bar{\Lambda}$. ii) The kinematic fit contributes with 1.7\%. iii) The model dependence of the detection efficiency is evaluated by changing the input $R$ with one standard deviation ($\pm~0.14$) in the {\sc ConExc} generator. This gives an uncertainty of 2.8\%. iv) The uncertainty of the integrated luminosity is 1.0\% \cite{lumi}. The individual uncertainties are assumed to be uncorrelated and are therefore added in quadrature, which yields a total systematic uncertainty of the cross section of 4.3\%. The systematic uncertainty in the effective form factor $|G|$ is obtained using error propagation and is half of the cross section.

In summary, the process $e^+e^-\to\Lambda\bar{\Lambda}$ ($\Lambda\to p\pi^-$, $\bar{\Lambda}\to\bar{p}\pi^+)$ is studied with 66.9 pb$^{-1}$ of data collected at 2.396 GeV. The cross section and the effective form factor are obtained to be $\sigma=118.7\pm5.3~(\rm stat.)\pm5.1~(\rm sys.)$ pb and $|G|=0.123\pm0.003~(\rm stat.)\pm0.003~(\rm sys.)$. The ratio $R=|G_E/G_M|$ is determined with unprecedented precision to be $R=0.96\pm0.14~(\rm stat.)\pm0.02~(\rm sys.)$. The relative phase between $G_E$ and $G_M$ is determined for the first time to be $\Delta\Phi=37^{\mathrm{o}}\pm12^{\mathrm{o}}~(\rm stat.)\pm6^{\mathrm{o}}~(\rm sys.)$. These results are obtained using the recent and most precise measurement of the asymmetry parameter $\alpha_{\Lambda}$. If, instead, the PDG value of $\alpha_{\Lambda}$ is used, the results become $R=0.94\pm0.16~(\rm stat.)\pm0.03~(\rm sys.)$ and $\Delta\Phi=42^{\mathrm{o}}\pm16^{\mathrm{o}}~(\rm stat.)\pm8^{\mathrm{o}}~(\rm sys.)$. 

The non-zero value of the relative phase implies that the EMFFs are complex at this energy. Hence, not only the $s$-wave but also the $d$-wave amplitude contribute to the production. Quantum number conservation in the one-photon exchange model only allows for $^3S_1$ and $^3D_1$ waves and their interference results in a polarized final state. This offers an unique and clean opportunity to learn about the $\Lambda\bar{\Lambda}$ interaction close to threshold. The prospects of this measurement have inspired the authors of Ref.~\cite{JU} to make predictions for the extracted observables in a recent theory paper. They used FSI potentials that were obtained from fits to data from the $\bar{p}p \to \Lambda\bar{\Lambda}$ reaction by the PS185 experiment at LEAR~\cite{PS185}. While the sensitivity of the energy dependence of the effective form factor $|G|$ of the $\Lambda\bar{\Lambda}$ FSI potential is very small, the predictions of $R$ and, even more, $\Delta\Phi$ depend significantly on the FSI potential. 
Our measurement slightly favors the Model I or Model II potential of Ref.~\cite{Haid1}. This illustrates the sensitivity of our data to the $\Lambda\bar{\Lambda}$ interaction.

The BESIII collaboration thanks the staff of BEPCII and the IHEP computing center for their strong support. This work is supported in part by National Key Basic Research Program of China under Contract No. 2015CB856700; National Natural Science Foundation of China (NSFC) under Contracts Nos. 11235011, 11335008, 11425524, 11625523, 11635010; the Chinese Academy of Sciences (CAS) Large-Scale Scientific Facility Program; the CAS Center for Excellence in Particle Physics (CCEPP); Joint Large-Scale Scientific Facility Funds of the NSFC and CAS under Contracts Nos. U1332201, U1532257, U1532258; CAS under Contracts Nos. KJCX2-YW-N29, KJCX2-YW-N45, QYZDJ-SSW-SLH003; 100 Talents Program of CAS; National 1000 Talents Program of China; INPAC and Shanghai Key Laboratory for Particle Physics and Cosmology; German Research Foundation DFG under Contracts Nos. Collaborative Research Center CRC 1044, FOR 2359; Istituto Nazionale di Fisica Nucleare, Italy; Joint Large-Scale Scientific Facility Funds of the NSFC and CAS; Koninklijke Nederlandse Akademie van Wetenschappen (KNAW) under Contract No. 530-4CDP03; Ministry of Development of Turkey under Contract No. DPT2006K-120470; National Natural Science Foundation of China (NSFC) under Contract No. 11505010; National Science and Technology fund; The Swedish Research Council; The Knut and Alice Wallenberg foundation, Sweden, Contract No. 2016.0157; U. S. Department of Energy under Contracts Nos. DE-FG02-05ER41374, DE-SC-0010118, DE-SC-0010504, DE-SC-0012069; University of Groningen (RuG) and the Helmholtzzentrum fuer Schwerionenforschung GmbH (GSI), Darmstadt; WCU Program of National Research Foundation of Korea under Contract No. R32-2008-000-10155-0.


\end{document}